\documentstyle[PASJadd,psfig]{PASJ95}

\newcommand{\vol}{}
\newcommand{\no}{}
\newcommand{\lett}{}
\newcommand{\spage}{}
\newcommand{\rdate}{1999 August 31}
\newcommand{\adate}{1999 November 9}

\begin{document}

\title{Deep X-Ray Observations of Supernova Remnants \\
       G~359.1$-$0.5 and G~359.0$-$0.9 with ASCA}
\author{Aya {\sc Bamba}, Jun {\sc Yokogawa}, Masaaki {\sc Sakano}, 
   and Katsuji {\sc Koyama}\thanks{CREST, Japan Science and 
   Technology Corporation (JST), 
   4-1-8 Honmachi, Kawaguchi, Saitama 332-0012.} \\
   {\it Department of Physics, Graduate School of Science, Kyoto University, 
   Sakyo-ku, Kyoto 606-8502} \\
   {\it E-mail(AB): bamba@cr.scphys.kyoto-u.ac.jp}}

\abst{
We present the results of deep ASCA observations 
of two shell-like radio supernova remnants (SNRs) 
located in the direction to the Galactic center (GC) region.  
Unlike the radio morphology,  G~359.1$-$0.5 shows center-filled X-rays 
with prominent K$\alpha$ lines of He-like silicon and H-like sulfur. 
The plasma requires at least two temperature  
components: a silicon-dominated cool plasma of 0.6~keV temperature and 
a sulfur-dominated higher temperature plasma of 4.4~keV.  
Because the absorption column is $\sim$ 6$\times 10^{22}$ H cm$^{-2}$, 
this SNR would be near to the GC.    
The spherical plasma is attributable to supernova ejecta with the total 
mass of Si and S being about $0.1$~$M_\odot$ 
and $0.3$~$M_\odot$, respectively. 
X-rays from G~359.0$-$0.9 trace the partial shell structure of 
the radio emission.  
The spectrum is well fitted to a single-temperature plasma of 0.4~keV 
with a non-solar abundance of magnesium or iron. 
Because the absorption column is not very large, 
$\sim$ 1.8$\times 10^{22}$~H~cm$^{-2}$, 
G~359.0$-$0.9 would be in front of the GC region.  
The total supernova energy, interstellar density near to the X-ray emitting 
shell and age of the SNR are estimated to be $1.2\times10^{51}$~erg,
0.5~cm$^{-3}$, and 1.8$\times10^4$~yr, respectively.  
We also discuss possible implications on the origin of the large-scale 
hot plasma surrounding the GC. 
}

\kword{Galaxy: center --- Galaxy: diffuse plasma --- ISM: abundances ---
 ISM: individual (G~359.1$-$0.5, G~359.0$-$0.9) --- supernova remnants
 --- X-rays: ISM}

\maketitle
\thispagestyle{headings}

\section{Introduction}

	Hard X-rays have opened a new window to see the Galactic center (GC) 
region.  
One of the remarkable discoveries is that a high-temperature ($\sim$ 10~keV) 
thin-thermal plasma is prevailing over an
$\sim$ 100 pc-radius region around the GC 
(Koyama et al.\ 1986, 1989, 1996; Yamauchi et al.\ 1990).  
The presence of the large-scale hot plasma suggests violent activities 
in the GC region in the past, although the origin is not yet clear.   
One of the relevant processes 
to produce such a high-temperature diffuse plasma 
would be multiple supernova explosions.  
In this context, X-ray observations of individual supernova remnants (SNRs) 
near to the GC may provide useful information not only on the SNR physics, 
but also on the origin of the GC plasma. 

	Since X-rays from SNRs are usually dominated in the low-energy band 
($\leq 2$ keV), 
and are easily absorbed by interstellar gas, an X-ray study of the SNRs near 
to the GC region is rather limited. 
High sensitivity and imaging capability in the hard X-ray band are 
required to detect SNRs behind a large absorbing medium, 
and to separate individual SNRs from the GC X-ray emissions,
a complex diffuse plasma, binary sources, and the other stellar objects.

	X-ray imaging spectroscopy also provides direct information on the 
nuclear synthesis, the total explosion energy and the age of SNRs, 
physical parameters of the surrounding interstellar environment, 
such as the density, its chemical compositions, and other related subjects: 
the star forming rate, the structure and the evolution 
in the central region of the Galaxy.

	The ASCA satellite, having high sensitivity 
in the hard X-ray band ($\geq 2$ keV) and high energy resolution, enables us 
to study more elaborate imaging spectroscopy 
than was possible with previous detectors.  
We conducted survey and pointing observations near to the GC region 
with the ASCA satellite, 
and found X-rays from radio SNRs, including new candidates.
Among them, this paper reports on the first detailed X-ray 
information and analyses of two radio SNRs, G~359.1$-$0.5 and G~359.0$-$0.9.

	G~359.1$-$0.5 was first identified as an SNR 
by a 4.9~GHz observation (Downes et al.\ 1979)
and by a 10.55~GHz observation (Sofue et al.\ 1984).  
Uchida et al.\ (1992b) found a shell-like structure at 1.4~GHz
surrounded by a $^{12}$CO ring.
Comparing the 21~cm 
absorption feature of the $^{12}$CO ring with the Galactic rotation curve, 
they concluded the location of this SNR to be near to the GC.
Although and Egger, Sun (1998) discovered  X-rays from  
G~359.1$-$0.5 with ROSAT,  the spectral parameters, 
such as the temperature and the chemical composition, 
are not well constrained, 
due to the poor statistics and limited energy resolution. 
Preliminary results of the ASCA observation on G~359.1$-$0.5 are found 
in Yokogawa et al.\ (1999).  

G~359.0$-$0.9 was first identified as an SNR by a 10.55~GHz observation
(Sofue et al.\ 1984) and by a 2.7~GHz observation (Reich et al.\ 1990),
and was later found to have an incomplete shell at the 843~MHz (Gray 1994).
Leahy (1989) first detected a partial shell of soft X-rays from G~359.0$-$0.9 
with the Einstein satellite, but no spectral information was reported.

	This paper presents more comprehensive ASCA results and 
analyses of these two SNRs. 
Particular care concerning the background subtraction 
was made to exclude any possible contamination of near-by 
bright X-ray sources and the GC plasma's contribution, 
of which the X-ray flux differs from position to position.

	We describe the observations and the method of data reduction in
section~2,  and the analyses in subsection~3.1 and
subsection~3.2, for G~359.1$-$0.5 and G~359.0$-$0.9, respectively.
Section~4 is devoted to results and discussions 
on the distances, chemical compositions and morphology of these SNRs, 
and also on a relevant subject, the origin of the GC plasma.

\section{Observations and Data Reduction}

\begin{figure}[hbtp]
  \psfig{file=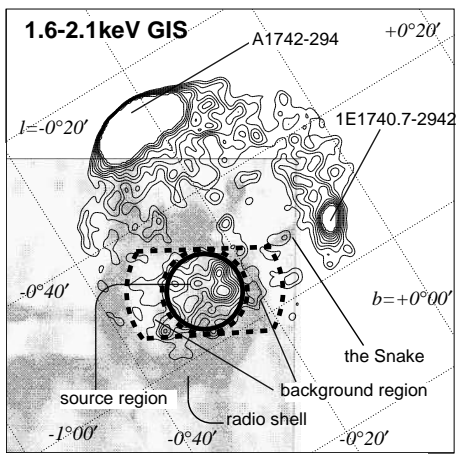,width=6cm}
  \caption{GIS contour map in the 1.6--2.1~keV band with Galactic coordinates, 
     superposed on the radio map by Gray (1994). 
     The contour levels are linearly spaced and are saturated at A~1742$-$294 
     and 1E~1740.7$-$2942.
     The source and background regions for spectral analyses are shown 
     with solid and dotted lines, respectively.
    }
  \label{G359.1$-$0.5_image}
\end{figure}


The GC region was observed with the X-ray astronomy satellite 
ASCA (Tanaka et al.\ 1994).  X-ray photons were collected with four XRTs 
(X-ray Telescopes; Serlemitsos et al.\ 1995) and simultaneously  detected 
with four detectors on the foci, 
which are two GISs (Gas Imaging Spectrometers; Ohashi et al.\ 1996) 
and two SIS (Solid-state Imaging Spectrometers; Burke et al.\ 1991) cameras.

During observation of the most prominent radio filament (the Snake) 
made on 1997 March 20--22, 
G~359.1$-$0.5 was located in the GIS field (see figure 1).  
However, it was only partially covered with the SIS field (see figure 3), 
in which 2-CCD chips for each SIS were operated with the faint or bright modes,
depending on the  high or medium bit rates.   
Since the 2-CCD data were significantly degraded
 by the Residual Dark Distribution (RDD) noise (T.Dotani et al.\ 1997, 
ASCA News 5, 14),  
we applied the RDD correction technique, 
which is only possible for faint-mode data.

G~359.0$-$0.9 was in the GIS field when the ASCA was pointing at 
the region of an unusual radio source the ``Mouse'' (Yusef-Zadeh, Bally 1987) 
and X-ray bursters SLX 1744$-$299/300 
on September 12--14 in 1998 (see figure 4).  
In this observation, because SIS was operated in the 1-CCD mode, 
G~359.0$-$0.9 was totally out of the SIS field.

In both observations, the GISs were always operated in the normal PH mode.
We excluded high-background data and non-X-ray events with the standard method
according to the user guide by NASA Goddard Space Flight Center (GSFC).
In total, the available exposures are  $\sim$ 80~ks of the GIS,  and
$\sim$35~ks of the SIS for G~359.1$-$0.5, and $\sim$~57 ks of GIS for 
G~359.0$-$0.9.

\section{Analyses}

\subsection{G~359.1$-$0.5 \label{sec:result:0.5}}

We found X-rays from the position of the radio SNR G~359.1$-$0.5 only 
in soft X-ray bands below about 3--4 keV.  
Figure~1 shows the GIS contour map in the energy of 1.6--2.1 keV,
in which band diffuse X-ray excess inside the radio shell
is most clearly seen.
In the hard X-ray band above 3 keV, on the other hand, we found 
no significant X-rays, 
neither along the radio shell nor in the center of the radio SNR. 
We also note that no significant X-rays were found from the Snake, 
the most prominent radio non-thermal filament (Uchida et al.\ 1992b).  

\begin{figure}[hbtp]
  \psfig{file=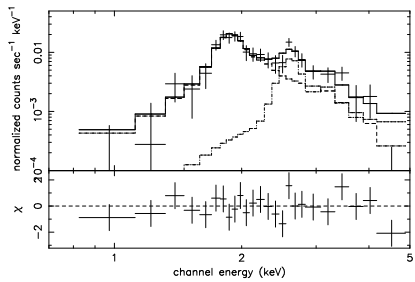,width=6cm}
  \caption{Background-subtracted GIS spectrum of G~359.1$-$0.5.
       The crosses are the data points. 
       The solid line represents the best-fit two-temperature 
       model, while the dotted lines are those of the individual components.} 
  \label{G359.1$-$0.5_spec}
\end{figure}


Since we found soft X-rays only from the center region surrounded 
by the radio shell, we made the GIS spectrum of the SNR center, 
accumulating X-ray photons from GIS 2 and 3 in the X-ray bright region of 
a 6$'$-radius circle, as shown by the solid line in figure~1.

As the background, we used the dotted line region in figure~1.
This background region was carefully selected so that 
(1) possible contamination from the two nearby bright sources, 
A~1742$-$294 and 1E~1740.7$-$2942, would be precisely  subtracted and 
(2) the contribution of the Galactic center plasma, distributed symmetrically 
around the GC, could be properly subtracted. 
For the two requirements, the angular distances of the background region from 
the GC and the other two bright sources were taken to be nearly 
the same as those of the source region.
The background-subtracted spectrum of GIS 2+3 is given in 
figure~2. 
We have checked the GIS energy scale by using the 1.86~keV Si line 
in the Galactic diffuse spectrum, 
since a long-term variation of a few percent level has been reported 
by Miyata (1996) and Yamauchi et al.\ (1999).
In this observation, we found that the GIS energy scale was very accurate 
with errors of less than 1\%. 

In the spectrum of G~359.1$-$0.5 (figure~2), 
we can notice two prominent emission lines at about 1.9 and 2.6 keV.  
To determine the accurate energy of the lines, we first fitted the spectrum to
a thermal bremsstrahlung (for continuum) and two Gaussian lines 
with an interstellar absorption.
The abundance for the interstellar gas was assumed to be solar, 
and the absorption cross sections were taken from Morrison and McCammon (1983) 
(hereafter, we refer to these absorption cross sections 
unless otherwise mentioned).

The best-fit center energies for the two lines 
were determined to be 1.86$^{+0.03}_{-0.04}$~keV and 
2.61$^{+0.07}_{-0.07}$~keV 
(here and after, the errors are 90\% confidence, 
unless otherwise mentioned).
These line-center energies are consistent with those of K$\alpha$ emission 
from helium-like silicon (He-like Si; 1.86~keV) 
and hydrogen-like sulfur (H-like S; 2.63~keV); 
hence, the observed line structures are attributable to these highly 
ionized atoms.
The best-fit line energies, on the other hand, are different 
from those of H-like Si (2.00~keV) and He-like S (2.45~keV). 

Although the presence of K-shell lines of He-like Si and H-like S
supports that 
the SNR X-rays are due to a thin thermal plasma, 
these two lines can hardly coexist in a single-temperature plasma,  
because ionization of sulfur atoms requires a higher temperature 
than that to ionize silicon atoms.  
In fact, we can reject any single-temperature plasma model, 
even though allowance is made for a non-ionization equilibrium 
or a non-solar abundance plasma.
Therefore, we applied a two-temperature model ({\tt MEKAL}:
the plasma code established by Mewe et al.\ (1985)
and Kaastra (1992)) with a  common interstellar absorption.  
The abundances of Si and S in each plasma were treated to be free parameters, 
whereas those of the other elements were fixed to the solar values 
(Anders, Grevesse 1989).

This model is statistically accepted within the 90\% confidence level.
The best-fit model and parameters are given in
figure~2 and table~1, respectively.
A remarkable result is that the sulfur abundance in the higher temperature 
plasma (here component 2) is larger than $\sim$40 of solar.

\begin{figure}[hbtp]
  \psfig{file=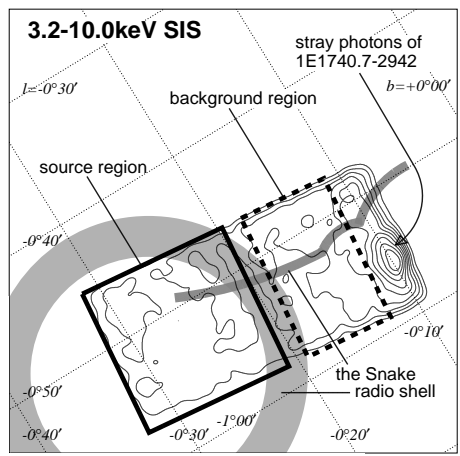,width=6cm}
  \caption{SIS contour map of G~359.1$-$0.5 in the 3.2--10.0~keV band 
    with Galactic coordinates, 
    superposed on a schematic diagram of the radio shell and the ``Snake.''
The contour levels are linearly spaced and are saturated 
at 1E~1740.7$-$2942.    
The source and background regions for spectral analyses are  
shown with solid and dotted lines, respectively.
    }
  \label{g3591_05_sis_image}
\end{figure}


Since SIS has a better energy resolution than GIS,  
a high-quality spectrum should be provided with SIS.  
However, as shown in figure~3, 
the SIS contour map in the 3.2--10.0~keV band, the relevant SIS field 
(2-CCD mode) covers only limited parts of the SNR.  
Furthermore, most of the SIS field is occupied by either G~359.1$-$0.5 
or 1E~1740.7$-$2942.  
Therefore, a proper background region, which is free from contamination 
of the bright source 1E~1740.7$-$2942 and contains the same diffuse 
GC emission as that in the source region, 
is not available in the present SIS field.  
Consequently, we concentrate on analyses of the emission-line structure, 
which is less sensitive to the background flux.  
Fortunately, we found that the contamination source, 1E~1740.7$-$2942, 
exhibits no significant emission line (Sakano et al.\ 1999a).  
We then selected the source and the background regions as indicated 
by the solid and dotted lines in figure~3.
As already noted, the surface brightness of the Galactic center 
diffuse X-rays decreases as the distance from the GC increases; 
hence, the simple background subtraction in the present case may cause an 
over-subtraction of the diffuse background.  
We therefore derived the iron-line fluxes from both the source 
and the background regions.  
The spectrum of the background region was subtracted 
from that in the source region, 
after normalizing the effective exposure by the iron-line flux ratio.  
These procedures are justified, 
because strong iron lines with a nearly constant equivalent width have been 
reported from the GC plasma (Koyama et al.\ 1989; Maeda 1998).        

To the background-subtracted spectrum, we fitted a model of 4 Gaussians and 
a power-law spectra with an absorption, of which the latter is simply 
a phenomenological model to represent the continuous spectrum.
The center energies of the Gaussian profiles were fixed to 
the theoretical values of 
He-like Si, H-like Si, He-like S, and H-like S.
The absorption column density was fixed to the best-fit value of 
the GIS analyses (see table~1).   
The best-fit fluxes, the ratios and most probable plasma temperatures 
under the assumption of ionization equilibrium are listed 
in table~2 (Mewe et al.\ 1985).

From this table, we can see that the plasma temperatures determined 
with Si and S are different from each other, 
and are consistent with those found in the GIS spectrum.  

To examine whether these two temperature plasmas have different spatial 
structures or not, 
we made the GIS and SIS images in the 1.6--2.1~keV and 2.1--3.2~keV bands; 
the former may represent the image of component 1, and the latter is for 
component 2.  
However, no significant difference between these two energy band images 
was found.

\subsection{G~359.0$-$0.9 \label{sec:result:0.9}}

	We made the  X-ray images in several energy bands and found that 
X-rays from G~359.0$-$0.9 were detected only below $\sim$ 3~keV.
The GIS contour map of G~359.0$-$0.9 in the soft X-ray band (0.7--1.5 keV) 
is shown in figure~4.  
Diffuse X-rays were clearly detected from the radio incomplete shell, 
although the X-ray shape is distorted by the bright X-ray sources 
SLX 1744$-$299/300.
For a spectral study, we selected the source and background regions as 
given in figure~4.  
The selection criteria of the background region are the same as 
those of G~359.1$-$0.5;  
the angular distances of the background region from the GC 
and from SLX 1744$-$299/300 are the same as those of the source region.   

\begin{figure}[hbtp]
     \psfig{file=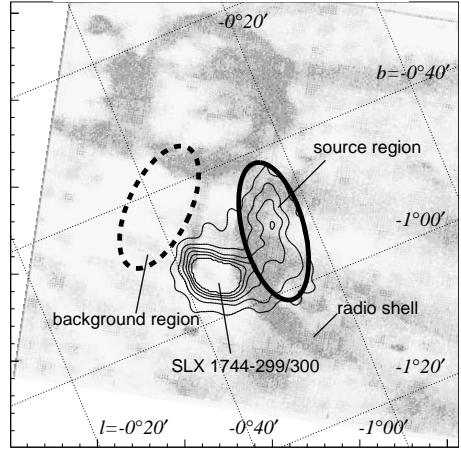,width=6cm}
 \caption{GIS contour map of G~359.0$-$0.9 in 0.7--1.5~keV 
      with Galactic coordinates superposed on the radio map (Gray 1994). 
      The contour levels are linearly spaced and are saturated 
      at SLX~1744$-$299 and SLX~1744$-$300.
      The source and background regions for spectral analyses are indicated by
      the solid and dotted lines, respectively.}
 \label{G359.0$-$0.9_image}
\end{figure}

 
We also examined any possible error of the GIS energy scale, 
as was done in the G~359.1$-$0.5 analyses, 
and found that the GIS energy is smaller than the proper value by about 2\%, 
and hence made a fine tuning of the energy scale to the spectrum.  
The background-subtracted (and energy-scale tuned) spectrum is shown 
in figure~5.
As expected from the multi-band X-ray images, 
most of the X-rays fall below $\sim$3~keV.
Because the spectrum exhibits  two clear  lines,
we applied a model of a thermal bremsstrahlung and two Gaussian profiles 
with an absorption, and determined the center energies to be 
1.36$^{+0.03}_{-0.03}$~keV and 1.86$^{+0.03}_{-0.03}$~keV.
Since these best-fit center energies are consistent with those of 
the K$\alpha$ lines  from He-like Mg at 1.35~keV 
and He-like Si at 1.86~keV,  
the origin of the X-rays is a thin thermal plasma.
We therefore applied a thin thermal plasma model, 
{\tt MEKAL}, with an absorption, 
and fixing the abundances of all elements to be solar [``model (a)'']. 
The best-fit model spectrum and parameters are shown 
in figure~5 and table~3, 
respectively.
\begin{figure}[htbp]
  \psfig{file=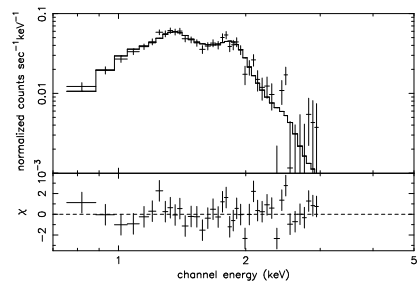,width=6cm}
  \caption{Background-subtracted GIS spectrum of G~359.0$-$0.9.
     The crosses and the solid line represent data points 
     and the best-fit model 
     [model (a): a model of thin thermal plasma with an absorption.
     All abundances are fixed to be solar one], respectively.}
  \label{G359.0$-$0.9_vmeka}
\end{figure}


This model is, however, rejected with a reduced chi-square of 49.8/38, 
leaving bump-like and dip-like residuals around 1.3~keV and 1.0~keV, 
respectively (figure~5).
Since these energies correspond to the emission lines of a He-like Mg 
and Fe-L line complex, 
we separately treated the abundances of Mg and Fe to be free and 
fitted the spectrum again [``model (b)'' and ``(c)''].
The best-fit spectra and parameters are given in 
figure~6, figure~7, and table~3 for models (b) and (c), respectively.
\begin{figure}[htbp]
  \psfig{file=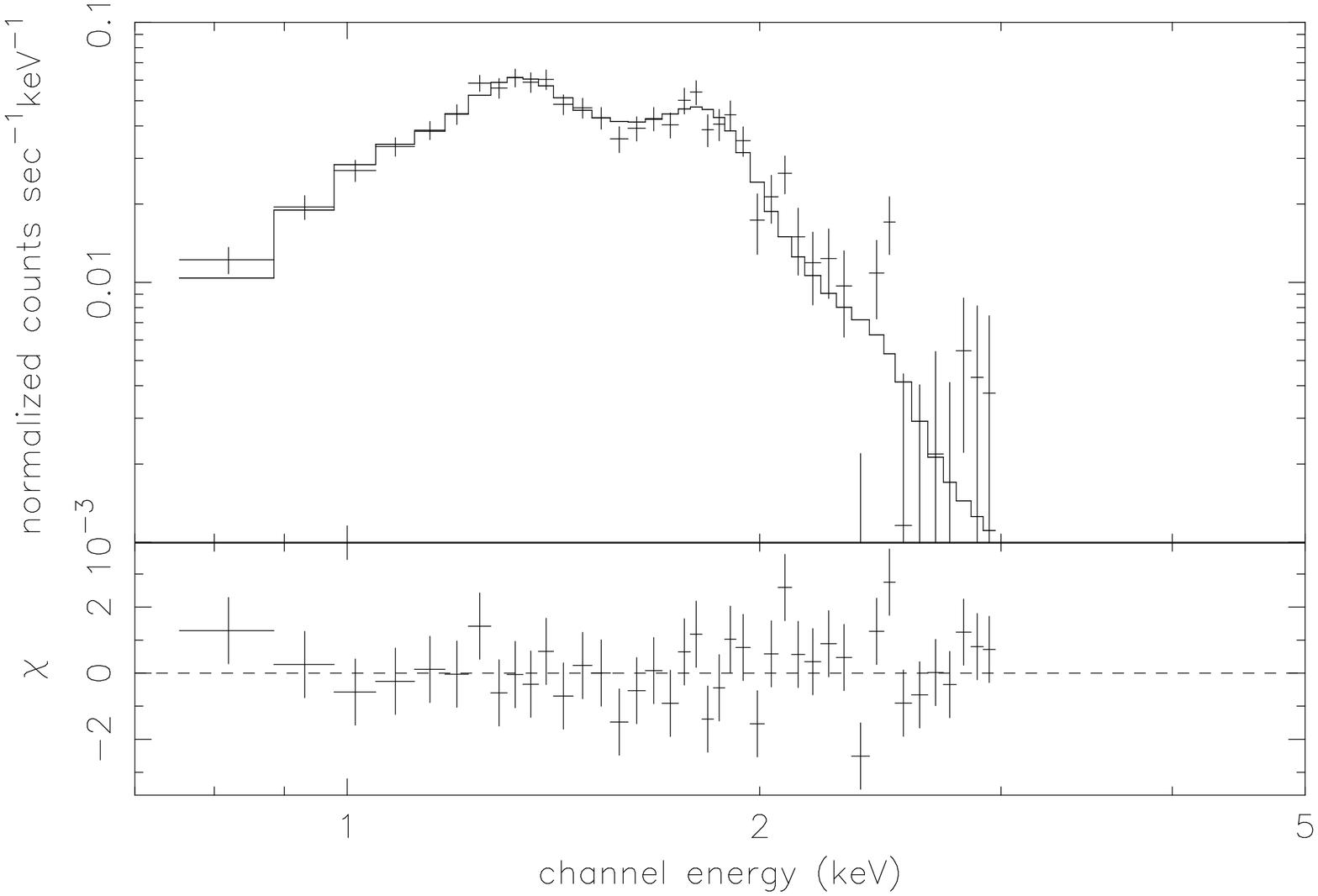,width=6cm}
  \caption{Same as figure~5, but for model (b):
     (a model of thin thermal plasma with an absorption.
     Same as model (a), but the abundance for Mg is treated to be free).}
  \label{G359.0$-$0.9_Mg}
\end{figure}
\begin{figure}[htbp]
  \psfig{file=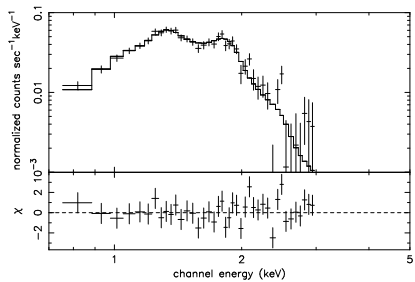,width=6cm}
  \caption{Same as figure~5, but for model (c):
     (a model of thin thermal plasma with an absorption.
     Same as model (a), but the abundance for Fe is treated to be free).}
  \label{G359.0$-$0.9_Fe}
\end{figure}

Although both the models (b) and (c) are acceptable, 
we further tried to fit the spectrum 
by treating the abundances of both Mg and Fe to be independent 
free parameters.  
However, the best-fit abundances have large errors,  
due to the limited statistics and the coupling of 
the He-like Mg line and Fe-L lines within the energy resolution of GIS.
Therefore, we conclude that the plasma of G~359.0$-$0.9 is non-solar abundance,
meaning that either Mg is over-abundant or  Fe is under-abundant. 

\section{Results and Discussions \label{sec:discuss}}
\subsection{G~359.1$-$0.5}

G~359.1$-$0.5 is found to exhibit a large absorption column of 
$\sim 5.9 \times 10^{22}$ H cm$^{-2}$.  
Since G~359.1$-$0.5 is reported to be surrounded by the $^{12}$CO ring 
for a total mass of about $2.5\times 10^6 M_{\odot}$ (Uchida et al.\ 1992b), 
local absorption due to the $^{12}$CO ring may not be ignored.
Assuming that the $^{12}$CO ring is a homogeneous shell with nearly 
the same shape of the G~359.1$-$0.5 radio shell,  
the absorption column due to the $^{12}$CO ring is estimated to be 
$\sim 3\times 10^{22}$~H~cm$^{-2}$.  
Therefore, we infer that the column density of the foreground interstellar 
matter is about $\sim 3\times 10^{22}$~H~cm$^{-2}$.
This value is equal to that of other X-ray sources near to the GC with the 
same Galactic coordinate of this SNR (Sakano et al \ 1999b); 
hence, this SNR would really be located near to the GC region with a distance
of about 8.5~kpc.
Thus, the diameter of the radio shell is estimated to be $\sim$ 57~pc, 
while that of the X-ray emitting central sphere is $\sim$ 28~pc.

We found that G~359.1$-$0.5 has at least two temperature plasmas;  
the cooler plasma (component 1) is abundant in Si, 
whereas the hotter one (component 2) is extremely over abundant in S.  
The center-filled thermal X-rays imply that these plasmas originated 
from the ejecta.
Assuming a  $\sim$28~pc-dimameter of spherical plasma of uniform density 
with a filing factor of 0.1,  
we estimate the total mass of Si and S to be about 
$0.1$~$M_\odot$  and $0.3$~$M_\odot$, respectively. 
However, no current theory of nucleosynthesis in supernova explosions 
predicts such a large mass of S compared to Si 
(see e.g. Thielemann et al.\ 1996).
This problem can be solved by assuming a smaller filling factor of S than
that of  Si; a smaller filling factor of S, less than 0.1, 
reduces the total mass of S to be acceptable for the model of 
Thielemann et al.\ (1996).

The context of the very small filling factor and the extreme richness of S 
lead us to suspect that the S-rich plasma is a ``shrapnel'' 
ejected from the massive progenitor of G~359.1$-$0.5, in analogy 
of the Vela SNR (Aschenbach et al.\ 1995).
However, the narrow band image including only the S-line (2.1--3.2 keV) shows 
no spatial structure like a ``shrapnel'', 
mainly due to a lack of photon statistics.

Uchida et al.\ (1992a) argued that the $^{12}$CO ring surrounding the SNR 
shell was created by stellar winds and/or multiple supernovae of O-type stars, 
and that several radio sources clustered at the center of the SNR are possibly
O-type stars.  
The X-ray spectrum of G~359.1$-$0.5 contains no clear emission line of Fe 
(see figure~2), 
and is thus consistent with the proposed O-star origin.
However, we could not quantitatively predict
on the mass of the progenitor because of a lack of photon statistics.

We found no shell-like X-ray from G~359.1$-$0.5, 
although the radio morphology shows an almost complete shell.  
Shell-like  X-rays may originate either by 
(1) a synchrotron mechanism in a shell, or  
(2) a thermal plasma made by the shock wave.

The lifetime of X-ray emission in case (1) depends on the magnetic field.  
Generally, the GC region is known to have a stronger magnetic field 
than the other regions of our Galaxy.
In particular, Robinson et al.\ (1996) observed Zeeman effects of 
three OH masers near to the shell of G~359.1$-$0.5, 
then directly estimated the magnetic field to be 0.4--0.6 mG, 
which is more than one order of magnitude larger than that in usual 
shell-likes SNRs ($\sim 10~\mu$G).  
High-energy electrons to emit synchrotron X-rays have a lifetime of 
$\sim 10^3$~yr in an $\sim 10~\mu$G field (e.g. Reynolds 1996);
hence, those in the shell of G~359.1$-$0.5 should be much shorter 
than $\sim 10^3$~yr.  
From the large diameter of the G~359.1$-$0.5 shell of 60~pc, 
this SNR would be middle age, or typically $\gtsim$ $10^4$~yr.  
Thus, no synchrotron X-ray from the shell would be expected from this SNR.

The evolved age of G~359.1$-$0.5 makes a shock-heated shell rather cool, 
with a typical temperature of less than a few 100~eV. 
Furthermore, if the shell interacts with the dense $^{12}$CO ring, 
the X-ray emitting shell density becomes larger, 
and hence the cooling time is much shorter.  
Therefore, X-rays should be very soft, and should be entirely absorbed 
by the large interstellar gas column.
Thus, the apparent lack of an X-ray shell of G~359.1$-$0.5 would be reasonable.

Rho and Petre (1998) proposed that SNRs with center-filled X-rays and 
a shell-like radio structure should be called 
``mixed morphology'' (MM) supernova remnants.
They show that the scale height of the MM SNRs' distribution from 
the Galactic plane is smaller than that of shell-like SNRs 
in both radio and X-ray bands, 
and that many MM SNRs are located near to molecular clouds or H~I clouds.
G~359.1$-$0.5 is located near to the Galactic plane, 
and is surrounded by the $^{12}$CO ring (Uchida et al.\ 1992b).
Thus, G~359.1$-$0.5 shares the common features of MM SNRs.

\subsection {G~359.0$-$0.9}

	The best-fit column density of $\sim 1.8\times 10^{22}$ H cm$^{-12}$ 
is smaller than that of the GC region 
(Sakano et al.\ 1999b). 
In fact, using the Galactic interstellar mass model 
by Olling and Merrifield (1998), and by Dame et al.\ (1987), 
we can estimate the distance to be 6 kpc.
Hence, G~359.0$-$0.9 would not be in the GC region, 
but would be a foreground SNR.   

	The shell-like morphology and a thin-thermal spectrum in X-rays 
suggest that the shell is a shock-heated plasma.  
The plasma, however, shows a partial shell  extending only 1/4 of 
a full circle.  
With the reasonable assumption 
that the supernova explosion was spherically symmetric, 
the apparent partial shell is attributable to an inhomogeneity of 
the interstellar medium; 
the direction to the X-ray emitting shell would have a higher density 
than the other directions, 
and would hence have a larger surface brightness than the others.  
For simplicity, we assume that the dense region covers $\pi /2$ str 
toward the partial shell; 
thus, 1/8 of the total explosion energy would be given in this direction.
Using the 6-kpc distance, the diameter and X-ray luminosity of the
partial shell are estimated to be 38 pc and $1\times 10^{34}$erg s$^{-1}$, 
respectively. 
Together with the observational temperature of 0.4~keV, 
we could solve the Sedov equation, and found that 
the total explosion energy ($E$), density ($n$) of the interstellar medium 
(dense region) and the age ($t$) are $1.2\times 10^{51}$~erg, 
0.5~cm$^{-3}$, and 1.8$\times10^4$~yr, respectively.
Because the total mass of the X-ray emitting shell is about 40$M_\odot$,  
most of them are attributable to the swept-up interstellar matter.  
Therefore, the result of the spectral fitting 
(in table~3) implies that the interstellar matter 
near to the G~359.0$-$0.9 is either over-abundant in Mg, 
or under-abundant in Fe.

\subsection {Comments on the Galactic Diffuse Plasma}

	The GC region is surrounded by the large scale hot plasma,
which emits fairly strong X-rays with many emission lines from He-like and 
H-like Si, S, Ar, Ca and Fe.   
The co-existence of highly ionized light elements (such as Si and S) and
heavier elements (such as Ca and Fe) implies that 
the plasma is not a single temperature. 
Kaneda (1996) and Kaneda et al.\ (1997) confirmed the two-temperature 
structure of the Galactic ridge plasma, 
which shows a very similar spectrum to that of the GC region. 
In fact, Maeda(1998) found that the GC plasma has two-temperature components. 

The low-temperature component ($\ltsim$1~keV) 
would be the same as that found with ROSAT (Snowden et al.\ 1997).
The plasma has a large-scale height of 1.9~kpc and a temperature of
0.3$-$0.4~keV.
Kaneda et al.\ (1997) suggested that the soft components of 
the Galactic ridge originated from a multiple supernova explosion.  
From the spectral similarity,
this scenario may be applied to the GC soft component.   
Then, can we find many individual SNRs which have a similar spectral shape?    
Because G~359.0$-$0.9 has a 0.4~keV temperature 
and strong emission line of Mg and Si, it is a possible candidate.  
The GIS flux at 0.5--10.0~keV of G~359.0$-$0.9 is $2.4\times 10^{-12}$~
erg~s$^{-1}$cm$^{-2}$, 
while that  of the GC plasma in a 25$^\prime$ diameter field is 
$8.3\times 10^{-12}$~erg~s$^{-1}$cm$^{-2}$ (Kaneda 1996).
We thus require about four G~359.0$-$0.9-like SNRs in this field.  
However, at this moment, the number density of the resolved X-ray SNRs or 
its candidates is far less than the requirement.

A more difficult issue is the origin of the high-temperature component.
The temperature of 10~keV and the size of 
about 100-pc diameter are inferred by the observation of Ginga and 
ASCA GC surveys (Koyama et al.\ 1989; 1996).
As far as we know, no SNR exhibits such a high temperature.  
In this sense, the high-temperature sulfur-rich plasma of 
G~359.1$-$0.5 is suggestive.  
If there are many  G~359.1$-$0.5-like SNRs, 
with an enriched abundance of not only S, but also other elements 
in high-temperature plasmas, 
the integrated emission may contribute, at least,  
some fractions  of the GC hot plasma.  
At present, we are still lacking any quantitative information of individual 
X-ray sources, like SNRs, near to the GC region. 
Thus, further systematic study is highly encouraged.\par
\vspace{1.5pc}\par
We would like to thank Dr.Y.\ Maeda for useful comments and discussion 
about G~359.1$-$0.5.
We also thank the members of the ASCA team.
This work was supported by the Research Fellowship of the Japan Society 
for the Promotion of Science for Young Scientists.

\small
\section*{References}
\re Anders E., Grevesse N.\ 1989, Geochim.\ Cosmochim.\ Acta 53, 197
\re Aschenbach B.,  Egger R., Tr\"umper J.\ 1995, Nature 373, 587
\re Burke B.E., Mountain R.W., Harrison D.C., Bautz M.W., Doty J.P., 
    Ricker G.R., Daniels P.J.\ 1991, IEEE Trans.\ ED-38, 1069
\re Dame T.M., Ungerechts H., Cohen R.S., de Geus E.J., Griener I.A., 
    May J., Murphy D.C., Nyman L.-\AA ., Thaddeus P.\ 1987, ApJ 322, 706
\re Dotani T., Yamashita A., Ezuka H., Takahashi K., Crew G., Mukai K., 
    the SIS team\ 1997, ASCA news 5, 14
\re Downes D., Goss W.M., Schwarz U.J., Wouterloot J.G.A.\ 1979, A\&AS 35, 1
\re Egger R., Sun X.\ 1998,
    The Local Bubble and Beyond, Lyman-Spitzer Colloquium, 
    Proceedings of the IAU Colloquium No.166, 
    (Springer-Verlag, Berlin), ed D.\ Breitschwerdt, M.J.\ Freyberg, 
    J.\ Tr\"umper Lecture Notes in Physics, 506, 417
\re Gray A.D.\ 1994, MNRAS 270, 835
\re Kaastra J.S.,\ 1992, An X-Ray Spectral Code for Optically Thin Plasmas
    version 2.0 (SRON, Leiden)
\re Kaneda H.\ 1996, PhD Thesis, The University of Tokyo
\re Kaneda H., Makishima K., Yamauchi S., Koyama K., Matsuzaki K., 
    Yamasaki N.Y.\ 1997, ApJ 491, 638
\re Koyama K., Awaki H., Kunieda H., Takano S., Tawara Y.\ 1989, Nature 339, 
    603
\re Koyama K., Ikeuchi S., Tomisaka K.\ 1986, PASJ 38, 503
\re Koyama K., Maeda Y., Sonobe T., Takeshima T., Tanaka Y., Yamauchi S.\ 
    1996, PASJ 48, 249
\re Leahy D.A.\ 1989, A\&A 216, 193
\re Maeda Y.\ 1998, PhD Thesis, Kyoto University
\re Miyata E.\ 1996, PhD Thesis, Osaka University
\re Mewe R., Gronenschild E.H.B.M., van den Oord G.H.J.\ 1985, A\&AS 
    62, 197
\re Morrison R., McCammon D.\ 1983, ApJ 270, 119
\re Ohashi T., Ebisawa K., Fukazawa Y., Hiyoshi K., Horii M., Ikebe Y., 
    Ikeda H., Inoue H. et al.\ 1996, PASJ 48, 157
\re Olling R.P., Merrifield M.R.\ 1998, MNRAS 297, 943
\re Reich W., F\"urst E., Reich P., Reif K.\ 1990, A\&AS 85, 633
\re Reynolds S.P.\ 1996, ApJ 459, L13
\re Rho J., Petre R.\ 1998, ApJ 503, L167
\re Robinson B., Yusef-Zadeh F., Roberts D.\ 1996, BAAS 28, 948
\re Sakano M., Imanishi K., Tsujimoto M., Koyama K., Maeda Y.\ 1999a, 
    ApJ 520, 316
\re Sakano M., Koyama K., Nishiuchi M., Yokogawa J., Maeda Y.\ 1999b, 
    Adv.\ Space Res.\ 23, 969
\re Serlemitsos P.J., Jalota L., Soong Y., Kunieda H., Tawara Y., Tsusaka Y., 
    Suzuki H., Sakima Y. et al.\ 1995, PASJ 47, 105
\re Snowden S.L., Egger R., Freyberg M.J., McCammon D., Plucinsky P.P., 
    Sanders W.T., Schmitt J.H.M.M., Tr\"umper J., Voges W.\ 1997, ApJ 485, 125
\re Sofue Y., Handa T., Nakai N., Hirabayashi H., Inoue M., Akabane K.\ 1984, 
    NRO report 33
\re Tanaka Y., Inoue H., Holt S.S.\ 1994, PASJ 46, L37
\re Thielemann F.-K., Nomoto K., Hashimoto M.\ 1996, ApJ 460, 408
\re Uchida K.I., Morris M., Bally J., Pound M., Yusef-Zadeh F.\ 1992a, 
    ApJ 398, 128
\re Uchida K.I., Morris M., Yusef-Zadeh F.\ 1992b, AJ 104, 1533
\re Yamauchi S., Kawada M., Koyama K., Kunieda H., Tawara Y., 
    Hatsukade I.\ 1990, ApJ 365, 532
\re Yamauchi S., Koyama K., Tomida H., Yokogawa J., Tamura K.\ 1999, 
    PASJ 51, 13
\re Yokogawa J., Sakano M., Koyama K., Yamauchi S.\ 2000,  
    Adv.\ Space Res.\ 25, 571
\re Yusef-Zadeh F., Bally J.\ 1987, Nature 330, 455

\begin{table*}[t]
 \begin{center}
Table.~1\hspace{4pt}Best-fit parameters for G~359.1$-$0.5 for the model of two
  thin thermal plasmas with an absorption.$^\ast$
 \end{center}
  \begin{tabular*}{\textwidth}{@{\hspace{\tabcolsep}
\extracolsep{\fill}}p{6pc}cccccc} \hline\hline\\[-6pt]
  & $kT$  & Si/H$^\dagger$ & S/H$^\dagger$ 
  & $N_{\rm H}$ 
  & Flux\ (0.7--10.0~keV) & $\chi^2$/d.o.f.$^{\ddagger}$ \\
Component  & (keV)  & & 
  & (10$^{22}$H cm$^{-2}$) 
  & (10$^{-13}$erg s$^{-1}$cm$^{-2}$) \\[4pt]\hline\\[-6pt]
1\dotfill & 0.6 (0.4--0.9) & 2.5 (1.2--8.4) & $<$0.9 & 5.9 (4.1--8.4) &
   7.8 & 17.18/17 \\ 
2\dotfill & 4.4 (2.2--13.2) & not determined & $>$ 38 & 5.9$^{\S}$ &
   6.1 & 
   $\cdots$ \\ \hline
  \end{tabular*}
 \vspace{6pt}\par\noindent
$^\ast$ Parentheses indicate 90\% confidence regions for one
 relevant parameter.
\par\noindent
$^\dagger$ Abundance ratio relative to the solar value.
\par\noindent
$^{\ddagger}$ d.o.f. $\equiv$ degree of freedom.
\par\noindent
$^{\S}$ Common with component 1.
\end{table*}

\begin{table*}[t]
 \begin{center}
Table~2.\hspace{4pt}Best-fit Si and S line fluxes  
  and most probable plasma temperature for G~359.1$-$0.5, 
  using the SIS data.$^\ast$
 \end{center}
\begin{tabular*}{\textwidth}{@{\hspace{\tabcolsep}
\extracolsep{\fill}}p{3pc}ccccccc} \hline\hline\\ [-6pt]
& He-like & H-like & H-like/He-like$^\ddagger$  
& $kT^{\S}$  \\
& (photons s$^{-1}$cm$^{-2}$) & (photons s$^{-1}$cm$^{-2}$) & &(keV) \\
[4pt]\hline\\[-6pt]
Si$^\dagger$\dotfill & $1.1\times 10^{-3}$ ($8.6\times 10^{-4}$--$1.3\times 10^{-3}$) & 
  $1.7\times 10^{-4}$ ($7.8\times 10^{-5}$--$2.7\times 10^{-4}$) & 
  0.16 (0.07 -- 0.25) & 0.5 (0.4 -- 0.5) \\
S$^\dagger$\dotfill & $1.5\times 10^{-5}$ ($ < 5.9\times 10^{-5}$) & 
  $5.3\times 10^{-5}$ ($1.7\times 10^{-5}$--$8.8\times 10^{-5}$) & 
  3.4 ($>$ 0.29)  & 1.7 ($>$ 0.8)  \\ \hline
\end{tabular*}
\vspace{6pt}\par\noindent
$^\ast$ Parentheses indicate 90\% confidence regions for one
 relevant parameter.
\par\indent
$^\dagger$\ Best-fit values and relevant parameters determined with silicon 
or sulfur.
\par\indent
$^\ddagger$\ Flux ratio between H-like and He-like K-shell lines.
\par\indent
$^{\S}$\ Plasma temperature determined with the line flux ratio under the assumption of ionization equilibrium.
\end{table*}

\begin{table*}[t]
 \begin{center}
Table~3.\hspace{4pt}Best-fit parameters for G~359.0$-$0.9 for 
thin thermal models.$^\ast$
 \end{center}
 \begin{tabular*}{\textwidth}{@{\hspace{\tabcolsep}
\extracolsep{\fill}}p{6pc}ccccccc} \hline\hline\\ [-6pt]
& $kT$ & Mg/H$^\dagger$ & Fe/H$^\dagger$ & $N_{\rm H}$ & Flux (0.7--10.0~keV) 
& $\chi^2$/d.o.f.$^{{\ddagger}}$\\
Model & (keV) & & & (10$^{22}$H cm$^{-2}$) & (10$^{-12}$erg s$^{-1}$cm$^{-2})$
& \\[4pt]\hline\\[-6pt]
(a)\dotfill & 0.3 (0.3--0.4) & $\cdots$ & $\cdots$ & 2.0 (1.9--2.2) & 2.3 
    & 49.84/38 \\ 
(b)\dotfill & 0.4 (0.3--0.5) & 1.7 (1.3--2.2) & $\cdots$ & 1.8 (1.5--2.0) &
      2.4 & 40.92/37 \\
(c)\dotfill & 0.4 (0.3--0.5) & $\cdots$ & 0.1 ($<$ 0.5) & 1.5 (1.1--1.8) & 
      2.4 & 41.18/37 \\ \hline
  \end{tabular*}
\vspace{6pt}\par\noindent
$^\ast$ Parentheses indicate 90\% confidence regions for one
 relevant parameter.
\par\noindent $^\dagger$\ Abundance ratio relative to the solar value.
\par\noindent $^{{\ddagger}}$\ d.o.f. $\equiv$ degree of freedom.
\end{table*}

\end{document}